# Smartphone based Driving Style Classification Using Features Made by Discrete Wavelet Transformation

Roya Lotfi, Mehdi Ghatee

Department of Computer Science, Amirkabir University of Technology, Tehran, IRAN.

URL: www.aut.ac.ir/ghatee

Emails: ghatee@aut.ac.ir

Abstract:

Smartphones consist of different sensors, which provide a platform for data acquisition in many scientific researches such as driving style identification systems. In the present paper, smartphone data are used to evaluate the driving styles based on maneuvers analysis. The data obtained for each maneuver is the speed of the vehicle steering and the vehicle's direct and lateral acceleration. To classify the drivers based on their driving style, machine-learning algorithms can be used on these data. However, these data usually contains more information than it is needed and cause a bad effect on the learning accuracy. In addition, they may transfer some wrong information to the learning algorithm. Thus, we used Haar discrete wavelet transformation to remove noise effects. Then, we get the discrete wavelet transformation with four levels from smartphone sensors data, which include low-to-high frequencies, respectively. The obtained features vector for each maneuver includes the raw signal variance as well as the variance of the wavelet transform components. On these vectors, we use the k-nearest neighbors algorithm for features selection. Then, we use SVM, RBF and MLP neural networks on these features to separate braking and dangerous speed maneuvers from the safe ones as well as dangerous turning, U-turn and lane-changing maneuvers. The results are very interesting.

1. Introduction

Driving style evaluation is one of the important problem in the field of traffic safety. In [5] the authors developed one of the best smartphone identification system in which the smartphone system is fixed on the dashboard of the car and collect the data from an accelerometer sensor, a gyroscope, a magnetometer to classify the driving styles. In [6] a smartphone placed in a fixed position in the vehicle is used to identify maneuvers. After the maneuver has been identified, a number of safe and dangerous maneuvers have been used to train the system. In addition to detect dangerous maneuvers, the driving style of individuals is named with dangerous and safe labels. Dai et al. [7] used a directional sensor, which has low accuracy at high speeds [8], to identify the position of smartphone axis relative to the car. They used this information to correct accelerometer data. In this research, smartphone can be located anywhere. Their purpose is merely to identify the behaviors that a drunk driver incurs, for example, changing the line of travel and the sudden movement.

Yu et al. [9] identified intelligent mobile sensors and developed a system for determining the hazardous driving events by combining front and rear camera data. It also showed that with the use of a front-facing camera and image processing algorithms, they can accurately identify 85% of the wearer's weariness, which is one of the main reasons for driving accidents. Some researchers also developed the smartphone-based system using the line identification system and image processing algorithms, they showed that their introduced system is good on low quality cameras.

Fazeen et al. [11] used smartphone accelerometer sensors that were stationary in the vehicle to detect safe driving of dangerous maneuvers such as a sudden change in line and sudden acceleration. They have been able to detect two sudden acceleration maneuvers and a sudden change of line, and inform the driver of their dangerous behavior.

The authors of [12] utilized smartphone accelerometer information to model how people drive. They designed a system for the environmental driving program. The purpose of these programs is to help drivers to optimize their fuel consumption. By knowing the driver of their dangerous habits and behaviors, the system helps them to drive better and thus less fuel. In [13] the authors have investigated the effectiveness of smart cellular sensors in identifying driving maneuvers. They compared the signals received from smartphone sensors to the signals received from the controller network and indicated that smartphones could be used to identify driving maneuvers. In [14], the data collected by Smart Positioning System (SIP) and the Controller Area Network were considered to warn the driver and to reduce fuel consumption.

Mohan et al. [15] used the accelerometer sensor, microphone, and integrated positioning system to find road conditions such as potholes and traffic conditions. In the system developed by Microsoft, the focus was not on driving style identification, however, for their analyses, they proposed a method for identifying braking and gas maneuvers using smartphones.

Li et al. [16] used an intelligent mobile accelerometer sensor to detect driving events, including identification of maneuvers and road conditions. They initially converted the acceleration received by the smartphone to the acceleration of the vehicle by using the alignment algorithm and identified maneuvers by placing the threshold on the modified accelerations. In [17] the authors used sensors data including accelerometer, and intelligent cellular positioning system at any desired location of the vehicle, and a rating mechanism for driver behavior evaluation.

Balcerek et al. [21] used intelligent mobile sensors and machine learning algorithms to detect braking maneuver, and showed that their system had a poorly correct and negative false positive rate.

In continuation of the these works, in this paper we consider smartphones which include two sensors: accelerometer and gyroscope. Also, the user has installed the smartphone as shown in Figure 1 on the car's steering wheel. This smart phone holder has been invented on the steering wheel by [3]. The second assumption is that the driver records his driving data at a rate of 20 Hz.

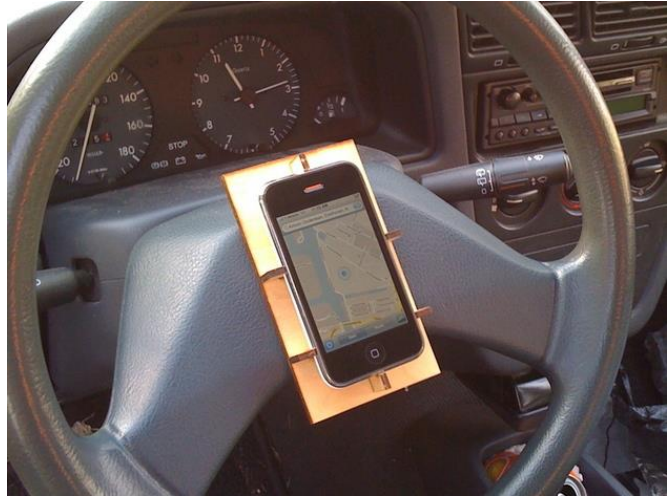

*Figure 1 Smartphone placed on the car's steering wheel*

The accelerometer sensor has three axes x, y, z, and represents the acceleration introduced into the smartphone as a three-component vector. Like that, the gyroscope sensor also has three components that show the angular velocity around each of its axes as a three-component vector. We use wavelet transformation [27] on the signals collected by these sensors and define a feature selection method on the wavelet features. We use different machine learning algorithms including SVM, MLP and RBF neural networks on the features and show that by a vey high accuracy it is possible to classify safe and dangerous maneuvers. This shows that we can evaluate the driving style for safety improvement.

2. Data preprocessing in driver evaluation system
    a. Wavelet transformation

This transformation can convert the signal into components of a different frequency, while preserving information related to the time domain. This conversion is obtained by passing a signal from a set of filters. In the first level decomposition, the signal passes through two high-pass and low-pass filters. The signal obtained after applying the high pass filter on the main signal is called the detail signal and the signal resulting from applying the low pass filter is called approximation. The following formula are the discrete wavelet transform at each level where g is the high pass filter and h is the low pass filter:

$$y_{low}[n] = \sum_{k=-\infty}^{+\infty} x[k].g[2n-k]$$

$$y_{high}[n] = \sum_{k=-\infty}^{+\infty} x[k].h[2n-k]$$

b. Wavelet transformation on smartphone sensors data

From smartphone sensors data, the speed of the vehicle's steering wheel, the vehicle's direct and lateral acceleration can be taken into account and then we can find the discrete wavelet transform of these data to four levels, which include low-to-high frequencies, respectively. The obtained feature vector for each maneuver includes the raw signal variance (before any filter), as well as the variance of the wavelet transform components. They can be represented as the following vector:

$$[dur, var(GZ), mean(GZ), var(A4), var(D4), var(D3), var(D2), var(D1), ...]$$

where dur is the time when a maneuver was carried out. There was an assumption that the maneuver would occur in less time is more dangerous than the maneuvering that occurs at a greater time. Therefore, the first component of the feature vector was considered as the start time to the end time of the maneuver. Then, we used the second and third component as the variance and mean speed of the vehicle's steering wheel (GZ), and eventually the variance of the approximation coefficients and the details of the discrete wavelet transform. Finally, the feature vector with 22 components was created. The components are featured in the following:

- Component 1: Runtime of maneuver
- Component 2: Variable speed of steering wheel
- Component 3: Average speed of steering wheel
- Component 4: Variance of coefficients of approximation A4
- Component 5: Variance of D4 detail coefficients
- Component 6: Variance of D3 detail coefficients
- Component 7: Variance of D2 coefficients
- Component 8: Variance of D1 detail coefficients
- Component 9: Variance of acceleration
- Component 10: Average of acceleration
- Component 11: Variance of A4 approximation coefficients
- Component 12: Variance of D4 detail coefficients
- Component 13: Variance of D3 detail coefficients (obtained from the lateral acceleration)
- Component 14: Variance of D2 detail coefficients (obtained from the lateral acceleration)
- Component 15: Variance of D1 detail coefficients (Obtained from the Acceleration)
- Component 16: Variance of direct acceleration Ax
- Component 17: Average of direct acceleration Ax
- Component 18: Variance of A4 approximation coefficients (obtained from direct acceleration)
- Component 19: Variance of D4 detail coefficients (obtained from direct acceleration)

- Component 20: Variance of D3 detail coefficients (obtained from direct acceleration)
- Component 21: Variance of D2 detail coefficients (Obtained from direct acceleration)
- Component 22: Variance of D1 detail coefficients (Obtained from direct acceleration)

Now, we need to select some features between these components for classification process.

    c. Feature selection phase

Feature selection methods are usually used as a pre-process phase of machine learning algorithms. Commonly, the descending gradient method is used in these methods to maximize the information obtained by a set of features. The weights of features indicate the relationship between the features and the output. The nearest neighbors algorithm is one of the best feature selection algorithm uses the descending gradient technique to maximize one-to-one validation. Since this method has no assumptions about the distribution of data and their scale, we use it for feature selection between the obtained components from wavelet transformation.

3. Maneuvers analysis in driver evaluation system
    a. Braking and gas maneuvers evaluation

For analyzing acceleration data related to braking and gas maneuvers within 3 seconds, the direct acceleration values recorded by the smartphone sensors are sampled. It is evident that the speed of the vehicle's steering and the vehicle's lateral acceleration has not any effect on how to perform brake and gas maneuvers, and only the sensed acceleration of the vehicle in these maneuvers should be considered. According to the collected data, if the acceleration changes are sampled in a 3-second window at about 1 m / s (0.11g), braking maneuver is very safe and if the acceleration changes for a maneuver is greater than 4 meters per second (0.45g), this maneuver is dangerous. When the acceleration interval is between these thresholds, the maneuver is between the safe and dangerous labels. However, we consider the range between 0.11g and 0.45g is also recognized as a safe range for maneuvering.

    b. Turning and U-turning maneuvers

For this kind of maneuvers, we use the feature selection o wavelet features. The first component, which gets the most weight is duration time of the maneuver. Clearly, this feature will characterize how the maneuver will be performed, as dangerous maneuvers are performed more quickly. The mean and the variance of the steering wheel, direct and lateral accelerations are weighted near zero, which confirms the initial claim that there are additional or inappropriate data in the raw signal. However, the variance of the coefficients of the velocity approximation of the vehicle's steering wheel, the variance of direct and lateral accelerations have high weights. Clearly, these data are more suitable for maneuvers analysis after the removal of the signal noise using a discrete wavelet transform. Selected features include those whose weight values are more than 0.1 are components 1, 4, 11, 18 and 22. These features are sent to MLP neural network by 2-6 hidden neurons. The results are given in Table 1.

*Table 1 MLP results on turning maneuvers classification*

| Number of hidden neurons | True-positive rate | Precision | AUC |
|---|---|---|---|
| 2 | 0.9541 | 0.9696 | 0.9453 |
| 3 | 0.9835 | 0.9882 | 0.9845 |
| 4 | 0.9917 | 0.9936 | 0.9909 |
| 5 | 0.9779 | 0.9796 | 0.9837 |
| 6 | 0.9993 | 0.9991 | 1 |

For the next method to classify safe and dangerous maneuvers, we use RBF neural network. First, K-mean algorithm computes the centers for the hidden layer neurons and then the network was randomly divided by 70% of the data for training, and 30% for the experiment, 100 times randomly. The comparison is presented in Table 2. The best answer was given with 6 hidden neurons.

*Table 2 RBF results on turning maneuvers classification*

| Number of hidden neurons | True-positive rate | Precision | AUC |
|---|---|---|---|
| 2 (one neuron for each class) | 0.9739 | 0.9663 | 1 |
| 4 (two neurons for each class) | 0.9777 | 0.9712 | 1 |
| 6 (three neurons for each class) | 0.9691 | 0.9683 | 0.9970 |

By SVM algorithm, we also find Precision=AUC=1.

    c. Lane-changing maneuvers

For this maneuver, we collected about 80 data with different conditions. By using feature selection it is shown that components 1, 11, 12 and 18 should be considered for classification of safe and dangerous lane-changing maneuvers. Here, duration time of maneuver has still a high impact on the classification of dangerous maneuvers from safety. We use again MLP, RBF and SVM classifiers for this maneuver. As an instance, the results of RBF network are presented in Table 3.

*Table 3 RBF results on lane-changing maneuvers classification*

| Number of hidden neurons | True-positive rate | Precision | AUC |
|---|---|---|---|
| 2 (one neuron for each class) | 0.9373 | 0.9394 | 0.9918 |
| 4 (two neurons for each class) | 0.9590 | 0.9607 | 0.9953 |
| 6 (three neurons for each class) | 0.9691 | 0.9683 | 0.9970 |

Also SVM classifies the lane-changing maneuvers by AUC=0.9707 and precision=0.9737.

d. An improvement by signal shapes

Now we wish to respond whether or not it is possible to define a new base functions for driving style evaluation. For this aim, we investigate the different signals collected in the different maneuvers. Our study shows that the response is positive. To show the value of such idea, we define a new problem to extract some new features for maneuvers classification. Our approach is to estimate any maneuver by a regression made by the summation of two Gaussian functions. Different evidences from different maneuvers can support this approach. For curve fitting in this part, we consider the data of the steering wheel angle signal and we estimate the parameters of the following function:

$$f(x) = a_1 \exp\frac{(x-b_1)^2}{2c_1^2} + a_2 \exp\frac{(x-b_2)^2}{2c_2^2}$$

After curve fitting, we use these parameters as new features for maneuvers classification. In Table 4, the performance derived from the implementation of different classification algorithms on these new features are presented for lane-changing maneuvers as an example. As one can see in this table, the new features provide almost similar results to the results given by traditional wavelet features. This shows that we can define special wavelet bases for driving evaluation.

*Table 4 Comparison between different learning algorithms on lane-changing maneuvers classification based on the wavelet features and new Gaussian function parameters*

| Number of hidden neurons | Considered features | True-positive rate | Precision | AUC |
|---|---|---|---|---|
| SVM | Wavelet features | 0.9737 | 0.9737 | 0.9707 |
|  | Gaussian function parameters | 0.9204 | 0.9356 | 0.9368 |
| RBF | Wavelet features | 0.9691 | 0.9683 | 0.9970 |
|  | Gaussian function parameters | 0.9119 | 0.9050 | 0.9716 |
| MLP | Wavelet features | 0.9993 | 0.9991 | 1 |
|  | Gaussian function parameters | 0.9825 | 0.9819 | 0.9832 |

4. Conclusion

In this paper, we focus on classification of driving maneuvers into safe and dangerous classes. We consider the collected data by smartphone sensors including acceleration, speed and steering wheel. Also the different types of learning algorithms were used to distinguish dangerous maneuvers from safe ones. Various criteria were used to examine the performance of learning algorithms and finally, the best answer for each of these algorithms was presented. We show that the features extracted by discrete wavelet transformation are very useful for maneuvers evaluation. Also we show that it is possible to define new bases for wavelet analysis to study on different signals collected by smartphones in different kinds of maneuvers. In the future works, one can try to find the best base for driving style evaluation by using wavelet transformation.